\documentclass[runningheads]{llncs}
\usepackage[T1]{fontenc}
\usepackage{graphicx}
\usepackage{multirow}
\usepackage{amsmath}
\usepackage{amssymb}
\usepackage{color}
\usepackage{hyperref}

\begin{document}
\title{MedCollab: IBIS-Guided Multi-Agent Collaboration with Hierarchical Disease Relation Chains for Clinical Diagnosis}
\titlerunning{MedCollab for Clinical Diagnostic Reasoning}
% If the paper title is too long for the running head, use an abbreviated title here.
%

\author{Yuqi Zhan\inst{1}$^*$ \and
Xinyue Wu\inst{1}$^*$ \and
Tianyu Lin\inst{1} \and
Yutong Bao\inst{1} \and
Xiaoyu Wang\inst{1} \and
Weihao Cheng\inst{1} \and
Huangwei Chen\inst{2}$^\dagger$ \and
Feiwei Qin\inst{1}$^\dagger$ \and
Zhu Zhu\inst{3}$^\dagger$}
\authorrunning{Y. Zhan et al.}
\institute{
Hangzhou Dianzi University \and
Zhejiang University \and
Children's Hospital, Zhejiang University School of Medicine,
National Clinical Research Center for Children and Adolescents' Health and Diseases\\
\email{huangwei\_chen@zju.edu.cn, qinfeiwei@hdu.edu.cn, zhuzhu\_cs@zju.edu.cn}\\[2pt]
{\small $^*$Equal contribution \quad $^\dagger$Corresponding author}
}
\maketitle              % typeset the header of the contribution

\begin{abstract}
Clinical diagnosis is a gradual process of evidence integration, in which physicians move from symptoms and medical history to examinations, competing hypotheses, disease relations, and treatment decisions. Large language models have advanced medical text understanding and generation. Yet their clinical use remains limited by weak evidence grounding, opaque reasoning, and inconsistent links among differential diagnosis, final diagnosis, diagnostic basis, and treatment planning. We introduce MedCollab, a multi-agent framework for full-cycle clinical diagnosis and report generation. MedCollab coordinates specialist and examination agents according to patient records. It structures agent deliberation with an Issue-Based Information System (IBIS) protocol, so that each diagnostic position is supported by patient-specific evidence and medical knowledge. It also builds Hierarchical Disease Relation Chains (HDRC) to connect accepted hypotheses through progression, complication, and comorbidity relations. During multi-round deliberation, a verifier-guided consensus module evaluates evidence support, medical plausibility, and logical conflicts. It then adjusts agent contributions and filters unsupported reasoning. Experiments on ClinicalBench and MIMIC-IV show that MedCollab outperforms leading LLMs and medical multi-agent baselines in diagnostic accuracy, evidence consistency, and clinical reasoning quality. These results indicate that structured and auditable collaboration can produce more faithful and clinically coherent diagnostic reports.

\keywords{Clinical NLP \and Multi-Agent \and Disease Relation Modeling}
\end{abstract}
\section{Introduction}
\setlength{\parskip}{0pt}
Clinical diagnosis is a full-cycle reasoning process rather than a single-step prediction task, encompassing differential diagnosis, preliminary diagnosis, final decision-making, and treatment planning. For clinical NLP, this process requires models to comprehend heterogeneous clinical texts, extract evidence from medical records, conduct multi-step reasoning, and generate coherent, complete, and evidence-faithful diagnostic reports~\cite {donroe2024clinical,choi2023using,zhu2026pathlens}. Despite the strong text understanding and generation abilities of LLMs, most only support isolated diagnosis prediction or partial report generation, lacking comprehensive coverage of the full diagnostic workflow from evidence interpretation and reasoning to treatment planning.

\begin{figure}[!t]
\setlength{\belowcaptionskip}{-6pt}
\includegraphics[width=\textwidth]{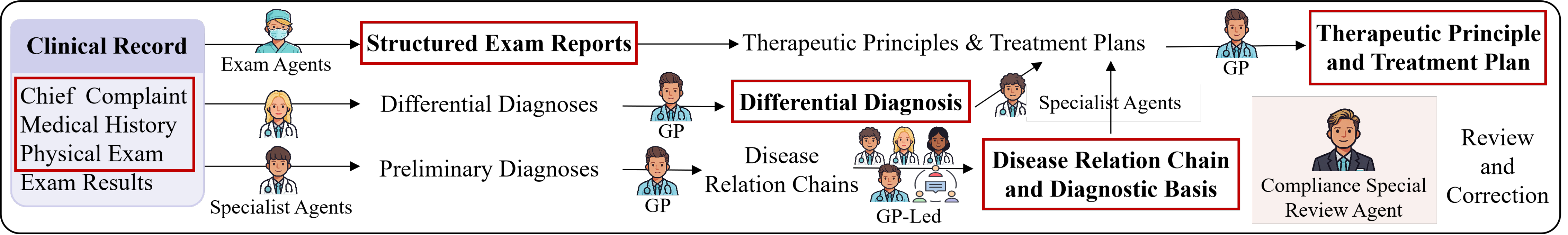}
\caption{Generation dependencies of the final diagnostic report in MedCollab. Key generated report sections are highlighted in red.}
\label{output}
\end{figure}

This problem is acute in online consultations with fragmented, unstructured medical data. Reliable diagnostic reports must identify possible diseases, ground them in patient evidence, and connect diagnoses to treatment. Existing LLM-based clinical systems often suffer from insufficient evidence grounding, incomplete transitions across diagnostic stages, and weak consistency among the diagnostic basis, differential diagnosis, and treatment plan~\cite{rutter2025importance,meyer2021patient}. %These issues further reduce the reliability and clinical usefulness of generated reports when multiple symptoms, overlapping disease manifestations, or comorbid conditions are involved.

To address these challenges, we propose MedCollab, an IBIS-guided multi-agent framework with Hierarchical Disease Relation Chains for full-cycle clinical diagnosis and report generation. It treats diagnosis as evidence-grounded clinical text reasoning and report generation, dynamically deploys agents based on patient records and fuses their outputs for report generation (Fig.~\ref{output}). Leveraging the IBIS protocol, MedCollab links each diagnostic hypothesis to traceable evidence~\cite{ibisconklin1988gibis,ibiskunz1970issues}, while HDRC organize accepted hypotheses into clinically meaningful progression and comorbidity relations, turning disconnected outputs into structured, evidence-grounded reports.

In summary, our contributions are as follows:
\begin{itemize}
    \item We formulate full-cycle clinical diagnosis as an evidence-grounded clinical NLP task. Unlike isolated disease prediction, this task requires models to connect differential diagnosis, preliminary diagnosis, final diagnosis, and treatment planning in a coherent report.
    \item We design MedCollab, an IBIS-guided multi-agent framework that constrains each diagnostic hypothesis with traceable evidence and structured clinical arguments.
    \item We introduce Hierarchical Disease Relation Chains to organize diagnostic hypotheses with disease progression and comorbidity relations, boosting logical consistency and evidence faithfulness of generated clinical reports.
    \item We evaluate MedCollab on ClinicalBench and MIMIC-IV. Results reveal it surpasses mainstream LLMs and medical multi-agent methods in diagnostic accuracy, evidence consistency, and clinical reasoning quality.
\end{itemize}

\section{Related Work}

Clinical LLMs have improved medical QA, diagnostic dialogue, and clinical text generation via biomedical pretraining, instruction tuning, and retrieval augmentation~\cite{wu2024pmcllama,singhal2025medpalm2,chen2024huatuogpto1medicalcomplexreasoning,xiong2024medrag}. AMIE further explores diagnostic conversation and differential diagnosis~\cite{mcduff2025amieDDx}. Yet most methods focus on isolated QA, dialogue, or single-step diagnosis, with limited modeling of consistency among diagnostic evidence, differential/final diagnosis, and treatment planning. This motivates a full-cycle diagnostic framework to compile multi-step clinical reasoning into consistent, evidence-based reports.

Medical multi-agent systems simulate collaborative clinical reasoning through role-specialized agents and multi-round discussion~\cite{tang2024medagents,kim2024mdagents}, department collaboration~\cite{yan2024clinicallab}, virtual multidisciplinary reasoning~\cite{chen_beyond_2026}, modular differential diagnosis~\cite{rose2025meddxagent}, tool-use and multimodal reasoning~\cite{li2024mmedagent}, and contradiction reduction~\cite{ma2025medla}. However, their intermediate reasoning is often free-form or task-specific, not unified and evidence-traceable. They also seldom model disease progression, complications, and comorbidities as explicit relations among hypotheses. These gaps motivate a more structured, clinically consistent framework.

\section{Methodology: MedCollab}

\begin{figure}[!t]
\setlength{\belowcaptionskip}{-6pt}
\includegraphics[width=\textwidth]{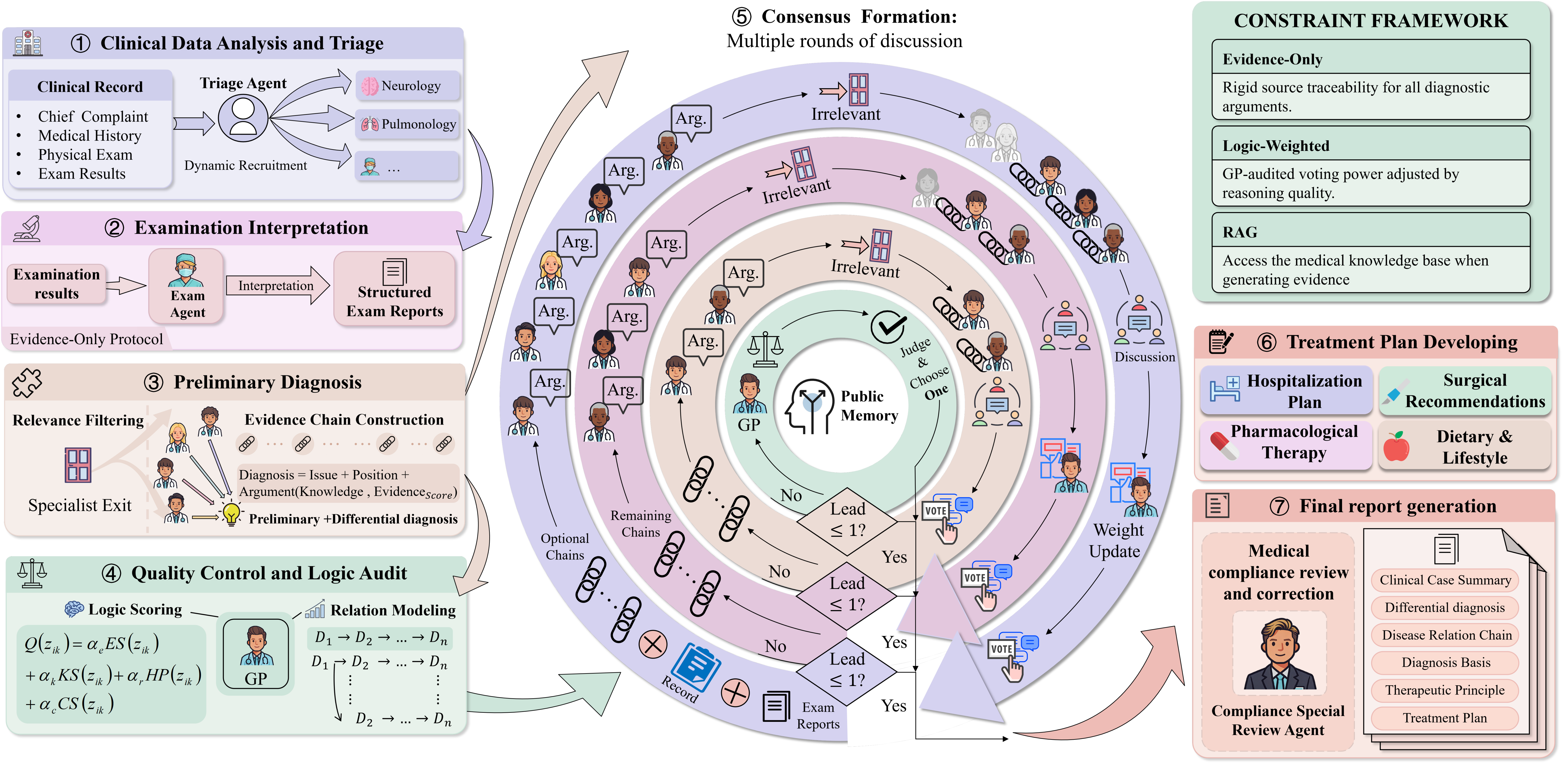}
\caption{The pipeline of MedCollab. The framework supports full-cycle clinical diagnosis through dynamic agent recruitment, IBIS-guided argumentation for traceable reasoning, and the construction of HDRC optimized by multi-round consensus.}
\label{fig1}
\end{figure}

\textbf {Overview.} As shown in Fig.~\ref {fig1}, the framework first dynamically recruits relevant specialist agents and exam agents (Sec.~\ref {sec:recruitment}). Active agents apply IBIS rules to standardize reasoning and produce reliable structured judgments (Sec.~\ref {sec:ibis}). These results are arranged into HDRC to resolve logical conflicts and reflect disease progression (Sec.~\ref {sec:HDRC}). Finally, a General Practitioner (GP)-led consensus module evaluates reasoning logic and adjusts agent weights to optimize the overall system (Sec.~\ref {sec:consensus}).

\subsection{Preliminaries: Dynamic Agent Recruitment}
\label{sec:recruitment}

Given patient case $S$ with complaint, history, physical exam, and raw findings, the GP agent performs coarse-grained triage to recruit two types of agents: specialist agents and exam agents. The specialist pool $\mathcal{A}_{spec}$ contains 23 common clinical department agents, while the exam-agent pool $\mathcal{A}_{exam}$ contains pathology, laboratory medicine, and radiology agents. Specialist agents are selected by relevance ranking, whereas exam agents are recruited according to the examination items available in $S$.

The GP agent assigns each specialist agent a relevance score $p(A_i \mid S) \in [0,1]$ based on $S$; higher $p(A_i \mid S)$ indicates a closer domain--case match. The top five specialist agents form the initial pool:
\begin{equation}
\mathcal{A}^{0}=\operatorname{Top}_{5, A_i \in \mathcal{A}_{spec}} \; p(A_i \mid S),
\end{equation}
where $\mathcal{A}_{spec}$ denotes the specialist-agent pool and $p(A_i \mid S)$ is the GP-estimated relevance of specialist agent $A_i$ to the patient case.

The recruited exam agents convert raw findings into structured exam reports, which are merged with the original record as evidence base $\mathcal{E}$. Each specialist agent then self-verifies: it remains active only if it identifies at least one in-domain condition supported by evidence spans in $\mathcal{E}$; otherwise, it exits. The remaining specialists form the active deliberation set:
\begin{equation}
\mathcal{A}^{\mathrm{act}}
=
\{A_i \in \mathcal{A}^{0} \mid \operatorname{Output}(A_i)\neq \emptyset\}.
\end{equation}
This two-stage design favors recall during initial triage while reducing irrelevant arguments in subsequent multi-agent deliberation.
\subsection{IBIS-Structured Argumentation for Diagnostic Reasoning}
\label{sec:ibis}

MedCollab represents each specialist's diagnostic output using the Issue-Based Information System (IBIS) protocol. For each active specialist agent $A_i \in \mathcal{A}^{\mathrm{act}}$, the diagnostic output is defined as a set of structured IBIS tuples:
\begin{equation}
\mathcal{D}_i=\{z_{ik}\}_{k=1}^{K_i},
\quad
z_{ik}=(I_{ik},P_{ik},\operatorname{Arg}_{ik},\mathcal{E}_{ik},\mathcal{R}_{ik}),
\end{equation}
where $K_i$ is the number of diagnostic hypotheses proposed by agent $A_i$.

\noindent\textbf{Issue ($I_{ik}$)} denotes the clinical question under investigation, such as what explains a patient's symptom, abnormal exam result, or disease progression.

\noindent\textbf{Position ($P_{ik}$)} denotes a proposed diagnostic hypothesis, such as ``rib fracture'' or ``hemothorax''.

\noindent\textbf{Argument ($\operatorname{Arg}_{ik}$)} provides the clinical rationale that connects the diagnostic hypothesis with patient-specific evidence and medical knowledge.

\noindent\textbf{Evidence ($\mathcal{E}_{ik}$)} is a subset of the evidence base $\mathcal{E}$, including traceable patient-specific evidence spans from the chief complaint, history, physical examination, laboratory results, imaging findings, or structured exam reports.

\noindent\textbf{Reference knowledge ($\mathcal{R}_{ik}$)} contains both parametric medical knowledge elicited from the backbone LLM and external medical evidence retrieved from PubMed. It is used to support the diagnostic argument and to check whether the proposed position is medically plausible.

A diagnostic position is considered valid by the verifier only when its key claims in $\operatorname{Arg}_{ik}$ are linked to at least one patient-specific evidence span in $\mathcal{E}_{ik}$ and one medical knowledge span in $\mathcal{R}_{ik}$.

\subsection{Hierarchical Disease Relation Chain Construction}
\label{sec:HDRC}

Flat diagnostic labels can identify possible diseases but often fail to express the relations among primary conditions, downstream complications, and co-existing diseases. MedCollab constructs a Hierarchical Disease Relation Chain.

Let $\mathcal{P}$ be the accepted diagnostic positions after evidence verification:
\begin{equation}
\mathcal{P}
=
\{P_{ik} \mid A_i \in \mathcal{A}^{\mathrm{act}},\;
z_{ik}\in \mathcal{D}_i,\;
Q(z_{ik}) \geq \tau_z\},
\end{equation}
where $Q(z_{ik})$ is the verifier score defined in Sec.~\ref{sec:consensus}, and we set $\tau_z$ to $0.7$.

HDRC uses two explicit relation operators: disease progression or downstream pathological relation $(\rightarrow)$, and comorbidity or co-existing disease relation $(+)$. HDRC candidate chains $\mathcal{H}$ are generated from the accepted diagnostic positions by constructing a candidate relation graph.  Candidate HDRCs are then enumerated from this relation graph.

A candidate disease relation chain $H_j$ is defined as:
\begin{equation}
H_j=d_1 \; r_1 \; d_2 \; r_2 \cdots r_{L_j-1} \; d_{L_j},
\quad
r_\ell \in \{\rightarrow,+\},
\end{equation}
where $d_\ell \in \mathcal{P}$ is a diagnostic position and $L_j$ is the number of diagnostic positions in the chain. For example, a trauma-related case may be represented as:
\begin{equation}
\text{Trauma} \rightarrow \text{Rib Fracture} \rightarrow \text{Hemothorax} \rightarrow \text{Anemia}.
\end{equation}

The operator $\rightarrow$ denotes a clinically plausible disease progression, complication, or downstream pathological relation supported by patient evidence and medical knowledge. The operator $+$ denotes comorbid or co-existing conditions that should be jointly considered in diagnosis and treatment planning.Multiple co-existing conditions can be represented in one HDRC through repeated $+$ operators.

The confidence score of a candidate HDRC is computed by aggregating verifier-supported relation evidence:
\begin{equation}
\operatorname{Score}(H_j)
=
\sum_{\ell=1}^{L_j-1}
\sum_{A_i \in \mathcal{A}_{d_\ell,d_{\ell+1},r_\ell}}
w_i \cdot q_{i,\ell,j},
\end{equation}
where $\mathcal{A}_{d_\ell,d_{\ell+1},r_\ell}$ is the set of agents that propose or support the relation $(d_\ell,d_{\ell+1},r_\ell)$, $w_i$ is the current weight of agent $A_i$, and $q_{i,\ell,j}$ is the verifier-supported relation-level quality score computed by the unified verifier scoring function in Sec.~\ref{sec:consensus}.

The final HDRC is selected as the highest-scoring chain under evidence coverage and contradiction constraints:
\begin{equation}
H^{*}=\arg\max_{H_j \in \mathcal{H}} \operatorname{Score}(H_j).
\end{equation}

\subsection{Verifier-Guided Consensus Optimization and Logic Auditing}
\label{sec:consensus}

MedCollab employs a verifier-guided consensus mechanism to optimize multi-agent diagnostic reasoning. The GP agent coordinates multi-round deliberation and aggregates verifier scores computed from evidence support, medical knowledge support, plausibility, and contradiction detection.

The verifier applies a unified scoring function to both diagnostic positions and HDRC relations:
\begin{equation}
\mathcal{V}(x)=
\frac{
\alpha_e ES(x)+\alpha_k KS(x)+\alpha_p PS(x)+\alpha_c(1-CS(x))
}{
\alpha_e+\alpha_k+\alpha_p+\alpha_c
}.
\end{equation}
where $x$ can be either an IBIS tuple or a candidate disease relation. $ES(x)$ measures patient-evidence support, $KS(x)$ measures medical-knowledge support, $PS(x)$ measures plausibility, and $CS(x)$ measures contradiction with patient evidence or medical knowledge. For IBIS tuples, $PS(x)$ denotes hypothesis plausibility; for HDRC relations, it denotes relation plausibility. The coefficients $\alpha_e$, $\alpha_k$, $\alpha_p$, and $\alpha_c$ regulate the relative contribution of these components, and all are initialized to 1. All component scores are assigned by the verifier agent under a fixed scoring rubric and normalized to $[0,1]$.

For each IBIS tuple $z_{ik}$, the diagnostic quality score is computed as:
\begin{equation}
Q(z_{ik}) = \mathcal{V}(z_{ik}).
\end{equation}

For each candidate HDRC relation $(d_\ell,d_{\ell+1},r_\ell)$ supported by agent $A_i$, the relation-level quality score is computed as:
\begin{equation}
q_{i,\ell,j}
=
\mathcal{V}(A_i,d_\ell,d_{\ell+1},r_\ell).
\end{equation}

For each active specialist agent $A_i \in \mathcal{A}^{\mathrm{act}}$, the logical inconsistency score at round $t$ is defined as:
\begin{equation}
\sigma_i^{(t)}
=
1-
\frac{1}{K_i}
\sum_{k=1}^{K_i}
\operatorname{clip}(Q(z_{ik}),0,1),
\end{equation}
where a higher $\sigma_i^{(t)}$ indicates weaker evidence grounding, poorer knowledge support, lower plausibility, or stronger contradiction.

All active agents are initialized with uniform weights $w_i^{(0)}=1$. Agent weights are updated through a normalized exponential penalty:
\begin{equation}
\tilde{w}_i^{(t+1)}
=
w_i^{(t)}\exp(-\lambda \sigma_i^{(t)}),
\quad
w_i^{(t+1)}
=
\frac{\tilde{w}_i^{(t+1)}}
{\sum_{A_m \in \mathcal{A}^{\mathrm{act}}} \tilde{w}_m^{(t+1)}},
\end{equation}
where $\lambda=0.8$ is the penalty coefficient. This update downweights agents with repeated unsupported or contradictory arguments while preserving those grounded in patient evidence and medical knowledge.

The consensus process iterates until the selected HDRC and final diagnosis remain unchanged for two consecutive rounds or the maximum number of deliberation rounds ($T=4$) is reached. The final report is generated from the accepted IBIS tuples and the selected HDRC, covering the diagnostic basis, differential diagnosis, final diagnosis, therapeutic principle, and treatment plan.
\section{Experiments and Results}
\subsection{Experimental Setup}

\begin{table}[b!]
\centering
\caption{Comparison of Diagnostic Precision on ClinicalBench (CB) and MIMIC-IV. Values are scaled by 100 (\%). Results are presented as \textbf{Value$_{CB}$ / Value$_{MIMIC}$}.}
\label{tab:diagnostic_results}
\fontsize{8pt}{9.6pt}\selectfont{
\begin{tabular}{l|cccc|c}
\hline
\textbf{Method} & \textbf{ACC }& \textbf{CDR }& \textbf{Entity-F1}& \textbf{DCA }& \textbf{$C_{dept}$}\\ \hline
\textit{Leading LLMs} & & & & & \\
HuatuoGPT-o1-8B~\cite{chen2024huatuogpto1medicalcomplexreasoning} & 42.9/32.6 & 32.7/23.9 & 29.1/26.6 & 76.7/59.8 & 80.8/82.5 \\
Baichuan4-Turbo~\cite{baichuan2024baichuan4} & 53.9/47.1 & 41.3/29.1 & 35.1/39.7 & 77.4/59.7 & 98.0/98.0 \\
GPT-4o~\cite{openai2024gpt4o} & 65.5/53.4 & 56.5/36.6 & 32.3/33.7 & 86.2/64.9 & 100.0/100.0 \\
Gemini-3-Flash~\cite{google2025gemini3} & 66.1/\underline{55.0} & 57.0/\underline{41.8} & \textbf{37.2}/36.3 & 87.6/\underline{71.6} & 100.0/100.0 \\
GLM-4.7~\cite{zhipuai2026glm47} & 66.9/51.6 & \underline{59.3}/37.6 & 30.5/32.8 & \underline{88.5}/68.4 & 100.0/100.0 \\
Qwen3-Max~\cite{yang2025qwen3} & 66.9/54.1 & 56.7/37.1 & 27.8/31.7 & 84.5/64.0 & 100.0/100.0 \\
Qwen3-8B~\cite{yang2025qwen3} & 52.9/44.9 & 39.6/28.6 & 31.8/\underline{37.8} & 72.6/65.5 & 97.6/98.1 
\\ \hline
\textit{Multi-Agents} & & & & & \\
MedLA~\cite{ma2025medla} & 63.4/49.6 & 54.3/35.0 & 32.1/35.8 & 85.7/64.7 & 99.9/100.0 \\
ClinicalAgent~\cite{yan2024clinicallab} & \underline{68.7}/51.6 & 51.3/29.2 & 25.6/28.8 & 75.6/54.3 & 100.0/100.0
\\ 
MEDDxAgent~\cite{rose2025meddxagent} & 36.9/31.8 & -/- & 26.8/26.0 & -/- & -/- \\ \hline
\textbf{MedCollab(Ours)}& \textbf{76.9/57.7} & \textbf{72.4/48.3} & \underline{36.7}/\textbf{37.8} & \textbf{94.3/83.7} & \textbf{100.0/100.0} \\ \hline
\end{tabular}
}
\end{table}
\begin{table}[!t]
\centering
\scriptsize
\caption{Clinician-involved Deepseek-V4-Pro-assisted evaluation of generated diagnostic reports. Results are presented as Value$_{\mathrm{CB}}$ / Value$_{\mathrm{MIMIC}}$.}
\label{tab:clinician_report_eval}
\fontsize{8pt}{10pt}\selectfont{
\setlength{\tabcolsep}{4.3pt}
\begin{tabular}{lcccc}
\hline
\textbf{Model} & \textbf{Fluency} & \textbf{Relevance} & \textbf{Completeness} & \textbf{Medical Correctness} \\ \hline
\textit{Leading LLMs} & & & & \\
HuatuoGPT-o1-8B~\cite{chen2024huatuogpto1medicalcomplexreasoning} & 4.60/4.91 & 3.25/3.30 & 2.60/2.91 & 2.40/2.55 \\
Baichuan4-Turbo~\cite{baichuan2024baichuan4} & 4.89/\underline{4.98}& 3.72/3.73 & 3.06/3.27 & 2.96/3.11 \\
GPT-4o~\cite{openai2024gpt4o} & \textbf{5.00}/\textbf{5.00} & 4.53/4.14 & 3.89/3.68 & 3.83/3.32 \\
Gemini-3-Flash~\cite{google2025gemini3} & 4.51/4.84 & 4.23/4.61 & 4.09/4.43 & 3.17/3.80 \\
GLM-4.7~\cite{zhipuai2026glm47} & \underline{4.96}/\textbf{5.00} & 4.36/4.66 & 4.04/4.27 & 3.77/4.05 \\
Qwen3-Max~\cite{yang2025qwen3} & \textbf{5.00}/\textbf{5.00} & 4.62/4.50 & 4.51/\underline{4.48} & 4.17/3.86 \\
Qwen3-8B~\cite{yang2025qwen3} & 4.94/4.89 & 3.57/3.43 & 3.19/3.18 & 2.79/2.57 \\
\hline
\textit{Multi-Agents} & & & & \\
MedLA~\cite{ma2025medla} & \textbf{5.00}/\textbf{5.00} & 4.51/4.52 & 4.17/4.14 & 3.96/4.02 \\
ClinicalAgent~\cite{yan2024clinicallab} & \textbf{5.00}/\textbf{5.00} & \underline{4.91}/\underline{4.82} & \underline{4.55}/4.16 & \underline{4.66}/\underline{4.41} \\
\hline
\textbf{MedCollab(Full)}& \textbf{5.00/5.00}& \textbf{4.93/4.87}& \textbf{4.67/4.63}& \textbf{4.81/4.66}\\
\hline
\end{tabular}
}
\end{table}

\textbf{Datasets and Data Processing.}
We evaluate MedCollab on two clinical datasets: (1) ClinicalBench~\cite{yan2024clinicallab}, which contains 1,500 bilingual English--Chinese clinical cases. This bilingual setting supports evaluation in Chinese clinical diagnostic scenarios while preserving cross-lingual comparability. (2) MIMIC-IV~\cite{johnson2023mimic}, from which we curated 983 cases. To ensure cross-dataset consistency, MIMIC-IV records were reformatted into a standardized clinical-case schema aligned with ClinicalBench. DeepSeek‑V4‑Pro was utilized as an auxiliary processing tool to map content to the target schema. After processing, a randomly sampled subset of the processed MIMIC‑IV cases was further reviewed by human clinicians for consistency. The construction and audit procedure is detailed in Appendix~\ref{app:data}.

\noindent\textbf{Evaluation Metrics.} We assess model performance across two primary dimensions: diagnostic precision and clinical reasoning quality.

For diagnostic precision, we evaluate triage routing using Department Classification Accuracy (DCA) and monitor hallucination suppression via Medical Factual Consistency (Entity-F1)~\cite{manning2008introduction}. The core diagnostic performance is measured by Accuracy (ACC), which assesses whether the primary diagnosis $p_{i}$ generated by the system matches the normalized ground truth $G_{i,1}$:
\begin{equation}
ACC = \frac{1}{N}\sum_{i=1}^{N}I(Norm(p_{i})=Norm(G_{i,1})),
\end{equation}
where $Norm(\cdot)$ denotes medical synonym normalization. To evaluate the framework's ability to simultaneously predict the correct department routing and disease diagnosis in complex cases involving multiple comorbidities, we also report Comprehensive Diagnostic Rate (CDR):
\begin{equation}
CDR = \frac{\sum_{i=1}^{N} S_{\text{guide}}(i) \times S_{\text{diagnosis}}(i)}{N},
\end{equation}
where $S_{\text{guide}}(i)$ and $S_{\text{diagnosis}}(i)$ are indicator functions for correctness of the $i$-th sample in the guide and diagnosis tasks, respectively, taking value 1 if the prediction is correct and 0 otherwise, and $N$ is the total number of test samples. We also report Department Consistency ($C_{dept}$), which measures whether the department predicted by the system matches a pre-defined valid department, serving as a complementary indicator of triage routing reliability. 

Moving beyond flat labels, we assess the logical coherence of generated reports across four structured sections: Diagnostic Basis (DB), Differential Diagnosis (DD), Therapeutic Principle (TP), and Treatment Plan (TX). Standard lexical overlap metrics such as BLEU and ROUGE-L are reported for textual similarity, but they often fail to capture medical synonyms or critical negations. We therefore prioritize RaTEScore~\cite{zhao2024ratescore}, an entity-aware semantic metric designed to assess causal logic and clinical truthfulness in reasoning chains.

\noindent\textbf{Multi-Agent Baseline Evaluation.} For agent baselines, we adapt each framework to our evaluation protocol. MEDDxAgent is evaluated using its top-1 extracted diagnosis and final DDx rationale, while MedLA is adapted from its original multiple-choice setting. Metrics unavailable under the original task design are reported as missing.

\noindent\textbf{Clinician-involved LLM-assisted Evaluation.} We evaluate generated reports with clinician and LLM assistance using a 1–5 rubric (Fluency, Relevance, Completeness, Medical Correctness) . We sampled up to 6 cases per department, yielding 144 ClinicalBench cases and 160 MIMIC-IV cases with anonymized model identities. For each department, two chief physicians independently rated two few-shot cases with criterion-level rationales; DeepSeek-V4-Pro then used these physician-provided rationales to score the remaining cases. 
\begin{figure}[t!]
\includegraphics[width=\textwidth]{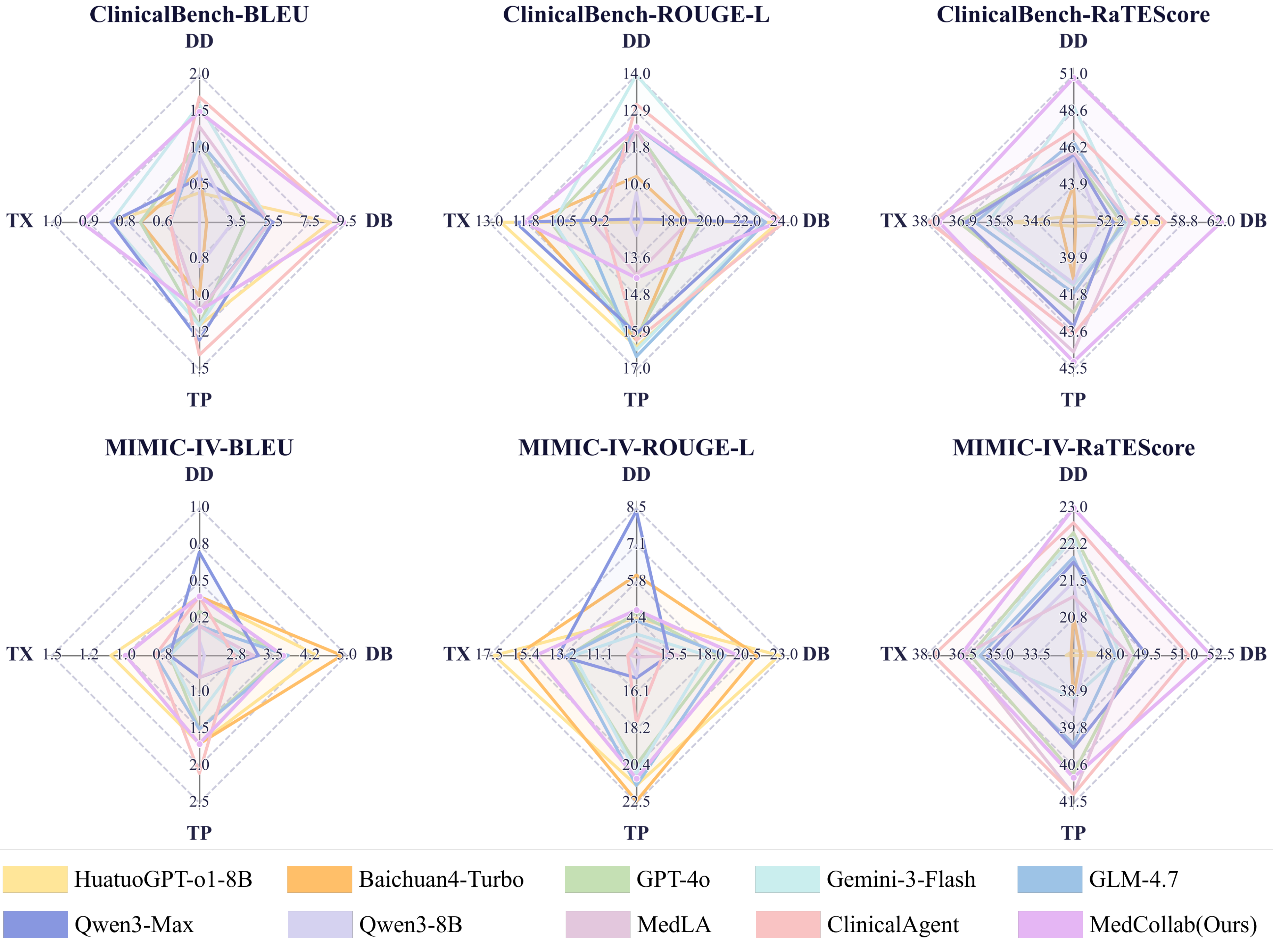}
\caption{Radar chart comparison of multidimensional report quality metrics (BLEU, ROUGE-L, and RaTEScore) across four clinical dimensions (DB, DD, TX, and TP) on ClinicalBench and MIMIC-IV. }
\label{fig:radar}
\end{figure}

\subsection{Main Results and Analysis}

We evaluate MedCollab against leading LLMs and specialized medical multi-agent systems across both datasets. For fair comparison, all multi-agent baselines and MedCollab are implemented with GPT-4o as the backbone LLM, while the GPT-4o baseline represents direct single-model inference. All methods use temperature 0, a maximum output length of 4096 tokens, the same evidence access, and consistent report specifications where applicable.

\noindent\textbf{Diagnostic Precision.} As shown in Table~\ref{tab:diagnostic_results}, MedCollab gains top ACC on ClinicalBench and MIMIC-IV, surpassing the best baselines by 8.2 and 2.7 percentage points respectively. It also achieves optimal CDR and DCA, reflecting better combined diagnosis-department judgment and stable agent allocation. MedCollab achieves competitive Entity-F1 while substantially improving ACC, CDR, and DCA, suggesting that IBIS-guided reasoning helps preserve entity-level factual consistency while improving diagnostic decision quality.

\noindent\textbf{Clinical Reasoning Quality.}
Fig.~\ref{fig:radar} shows that MedCollab achieves strong RaTEScore performance across diagnostic basis, differential diagnosis, therapeutic principle, and treatment plan, especially on evidence-sensitive reasoning dimensions. Although some LLMs obtain competitive BLEU scores, their lower RaTEScore indicates that surface overlap does not necessarily reflect clinically faithful reasoning. Table~\ref{tab:clinician_report_eval} further confirms that MedCollab achieves the best overall relevance, completeness, and medical correctness under clinician-involved LLM-assisted evaluation.

\begin{figure}[!b]
\includegraphics[width=\textwidth]{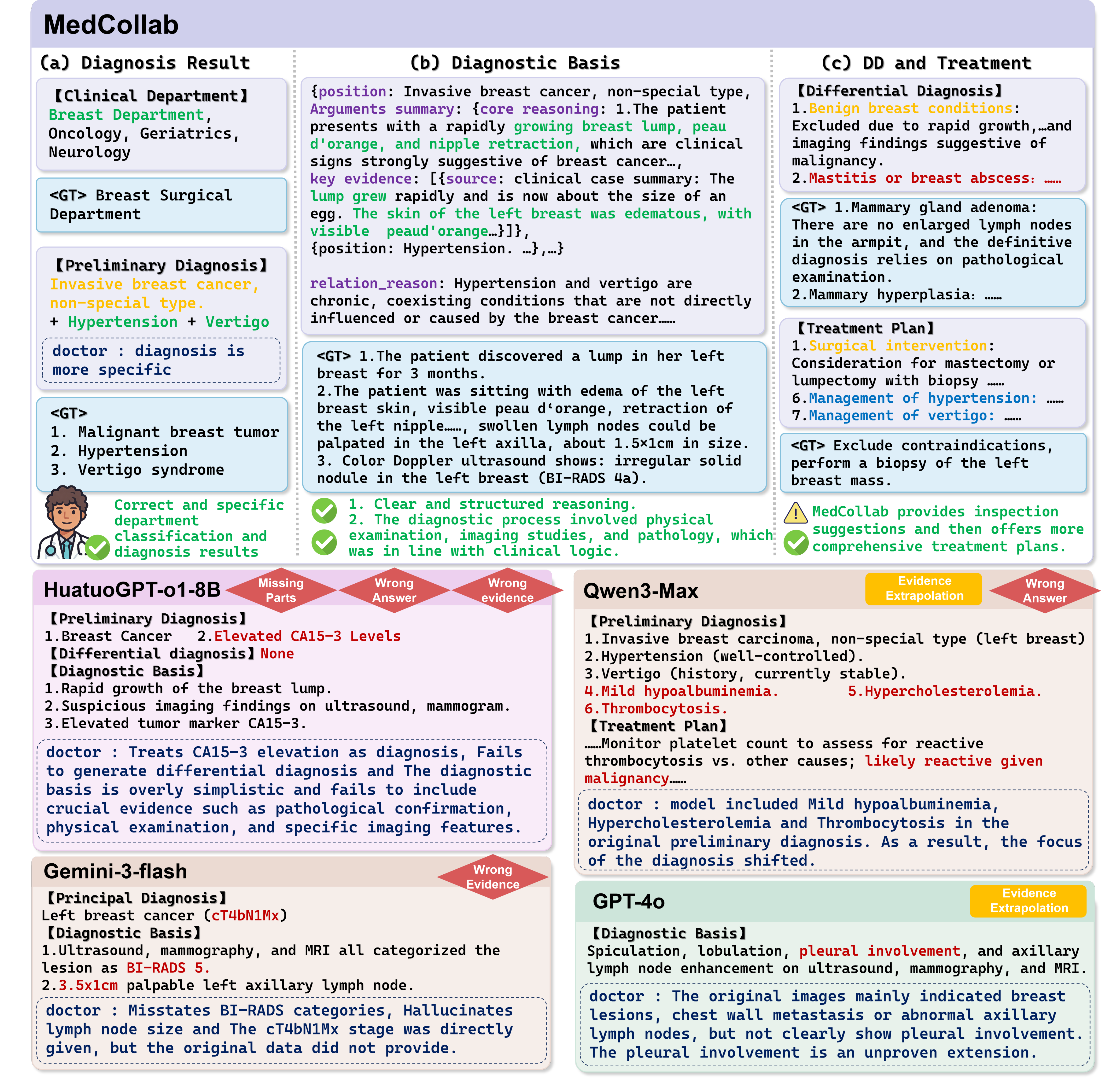}
\caption{A case study on the generated diagnosis report. Green indicates content that is consistent with the ground truth, red indicates errors, yellow indicates more detailed classifications accepted by clinicians, and blue indicates supplementary information recognized by clinicians, but this information is not included in the ground truth.}
\label{fig:output}
\end{figure}

After quantitative assessment, Fig.~\ref{fig:output} displays a qualitative case of diagnostic report generation. In the case study, MedCollab can generate standardized reports matching reference results and provides additional treatment suggestions that were marked as acceptable by clinicians during case review.

\subsection{Ablation Study}

To evaluate the contribution of each core component, we conducted an ablation 
study on ClinicalBench and MIMIC-IV, with results presented in Table~\ref{tab:ablation}.
\begin{table}[!t]
\centering
\caption{Ablation study on the ClinicalBench and MIMIC-IV dataset. Results are presented as Value$_{CB}$ / Value$_{MIMIC}$.}
\label{tab:ablation}
\fontsize{8pt}{10pt}\selectfont{
\setlength{\tabcolsep}{4.3pt} 
\begin{tabular}{lccccl}
\hline
\textbf{Method} & \textbf{ACC}& \textbf{Entity-F1}& \textbf{DB RaTE}& \textbf{DD RaTE}& \textbf{TX RaTE}\\
\hline
w/o Logic Auditing & 49.7/41.5& 29.6/32.9& 51.7/46.8& 46.6/22.0& 37.5/34.9\\
w/o HDRC& 52.9/40.2& 31.0/32.6& 51.7/45.3& 46.2/21.9& 37.0/35.0\\
\textbf{MedCollab (Full)} & \textbf{76.9/57.7}& \textbf{36.7/37.8}& \textbf{62.0/52.1}& \textbf{50.8/23.0}& \textbf{37.6/36.6}\\
\hline
\end{tabular}
}
\end{table}
\noindent\textbf{Impact of Logic Auditing.} Removing the Logic Auditing mechanism yields the largest degradation, with ACC dropping from 76.9\% to 49.7\% on ClinicalBench, suggesting that verifier-guided logic auditing contributes to more consistent inter-specialist reasoning.

\noindent\textbf{Impact of Relation Chain.} Removing HDRC reduces ACC from 76.9\% to 52.9\% and DB RaTEScore from 62.0\% to 51.7\%, suggesting that relation modeling contributes to diagnostic coherence.

\section{Conclusion}

We presented MedCollab, an IBIS-guided multi-agent framework for full-cycle clinical diagnosis and diagnostic report generation. By combining dynamic specialist recruitment, evidence-traceable IBIS argumentation, Hierarchical Disease Relation Chains, and verifier-guided logic auditing, it transforms fragmented diagnostic predictions into structured and pathology-aware clinical reasoning. Experiments on ClinicalBench and MIMIC-IV show that MedCollab consistently improves diagnostic accuracy, evidence consistency, and clinical reasoning quality over leading LLMs and medical multi-agent baselines. Ablation results further suggest that both Logic Auditing and HDRC contribute to coherent disease-relation reasoning. The bilingual English--Chinese evaluation also makes the benchmark setting more aligned with Chinese clinical NLP scenarios while preserving cross-lingual comparability.

\begin{credits}
\subsubsection{\ackname}
We gratefully acknowledge support from the National Key Research and Development Program of China (No. 2023YFC2706400).

\subsubsection{\discintname}
The authors declare no competing interests.
\end{credits}

% References will then be sorted and formatted in the correct LNCS style.
%
\bibliographystyle{splncs04}
\bibliography{NLPCC/refs1}

\appendix
\section{Additional Evaluation Details}

\subsection{Dataset Construction}
\label{app:data}

DeepSeek-V4-Pro was used only to map source content into the unified schema; its prompt explicitly prohibited rewriting, completion, inference, and unsupported additions. Two clinical physicians reviewed a department-stratified random sample of 10\% of cases, with at least five cases per department. No prohibited additions were identified in the reviewed samples.

\begin{figure}
\centering
\includegraphics[width=\textwidth]{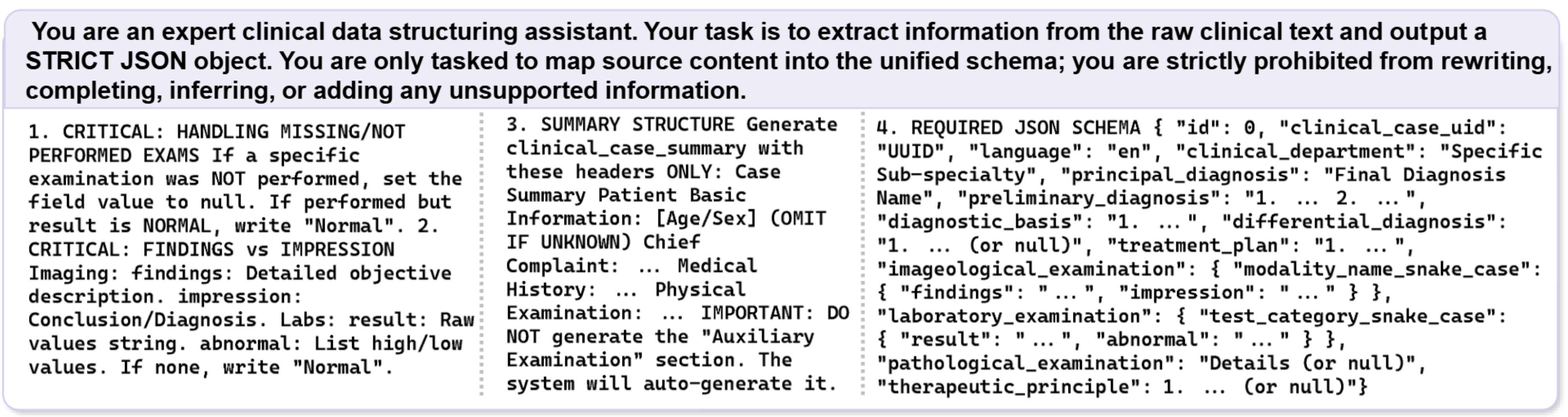}
\caption{Prompt used for DeepSeek-V4-Pro-based clinical case schema mapping.}
\label{fig:eva}
\end{figure}

\end{document}